\documentclass{aa}
\usepackage{graphicx}
\usepackage{graphics}
\usepackage{amssymb}                                                             
\usepackage{latexsym}
\begin{document}
\title{The solar neighbourhood age-metallicity relation - \\ 
does it exist?\thanks{Based on results from the ESA Hipparcos
satellite}}


   \author{Sofia Feltzing
          \inst{1}
          \and Johan Holmberg \inst{1}
	  \and Jarrod R. Hurley \inst{2}
          }

   \offprints{Sofia Feltzing}

   \institute{Lund Observatory, Box 43, SE-221 00 Lund, Sweden\\
              \email{sofia, johan@astro.lu.se}
         \and
             AMNH Department of Astrophysics, 79th Street at Central Park West 
     New York, NY, 10024-5192, USA \\
             \email{jhurley@amnh.org}
             }

   \date{}

\abstract{We derive stellar ages, from  evolutionary tracks,
and metallicities, from Str\"omgren photometry, for a sample
of  5828 dwarf and
sub-dwarf stars from the Hipparcos Catalogue. This
stellar disk sample is used to
investigate the age-metallicity diagram in the solar
neighbourhood. Such diagrams are often used to derive a so called
age-metallicity relation. Because of the size of our
sample, we are able to quantify the impact on such diagrams, and
derived relations, due to different selection effects.  Some of these
effects are of a more  subtle sort, giving rise to erroneous
conclusions. In particular  we show that [1] the age-metallicity
diagram is well populated at all ages and  especially that old, metal-rich
stars do exist,  [2] the scatter in metallicity at any given age is
larger than the  observational errors, [3] the exclusion of
cooler dwarf stars  from an age-metallicity sample
preferentially excludes old, metal-rich stars, depleting the upper
right-hand corner  of the age-metallicity diagram, [4] the distance
dependence found in the Edvardsson et al. sample by Garnett \&
Kobulnicky is an  expected artifact due to the construction of the
original sample.  We conclude that, although some of it can be
attributed to  stellar migration in the galactic 
disk, a large part of the observed scatter
is intrinsic to the formation processes of stars. 
} \maketitle

%

\section{Introduction}

Ages and metallicities of dwarf stars in the solar neighbourhood
provide a unique record of the progressive chemical enrichment of the
interstellar medium where the stars formed. As time progresses the
interstellar medium  becomes more and more enriched in
heavy elements. Thus we may expect  the more recently formed stars to
have a higher metallicity than those  formed at an earlier epoch.
By obtaining ages as well as metallicities for a 
representative stellar sample we should therefore be able to derive an
age-metallicity relation. Such a relation would provide a strong constraint
on any model of galactic chemical evolution. The picture might of course
be more complicated, e.g. including infall of unprocessed gas or stellar
migration. However, such processes can be modeled and the
observed age-metallicity diagram would again provide a strong constraint
on any model.

Are such simplistic assumptions borne out by observational evidence?
The current evidence, from both new and older studies, appear
to point in conflicting directions. The studies by Rocha-Pinto et al.
(2000) and Twarog (1980ab) do find good correlations
between age and metallicity for dwarf stars in the local galactic disk.
On the other hand several 
recent investigations appear to indicate that in fact the picture
is rather more complicated with metal-rich stars being both young and
old, e.g. Edvardsson et al. (1993), Carraro et al. (1998),
 Chen et al. (2000), and Feltzing \& 
Gonzalez (2001). For example in Edvardsson et al. (1993) the derived
age-metallicity relation has a scatter around the mean [Fe/H] at
a given age that is four times as large as the relative error on
individual [Fe/H] values. Thus strongly indicating that the observed scatter
is indeed real and not due to observational errors. As pointed out by 
Carraro et al. (1998) this means that the challenge for models
of galactic chemical evolution has shifted from focusing on trying to
reproduce an average trend, produced by binning in age and calculating
mean [Me/H] (the age-metallicity relation), to reproduce the overall
trends {\sl as well as} the observed scatter.

 However, it should
be remembered that most of the studies mentioned above include small numbers
of stars, ranging from a few (Feltzing \& Gonzalez 2001) to $\sim 200$
Edvardsson et al. (1993). Several of them  are 
 concerned mainly with stellar abundance analysis which
naturally limited the stellar samples in size. Furthermore, these samples
were selected in order to study specific aspects of the galactic chemical
evolution (Edvardsson et al. 1993 and Chen et al. 2000) or a specific
type of stars (Feltzing \& Gonzalez 2001) and are therefore not directly
representative of the solar neighbourhood. 

The new large database of good stellar parallaxes presented in 
the Hipparcos Catalogue (ESA 1997) provides a unique opportunity to 
investigate the age-metallicity plot using a larger number of stars. 
In the study presented here we investigate, using 5828 stars, the
structures and trends in the age-metallicity plot.

The article is organized as follows: in Sect. 2 we describe the stellar
sample used and how we determine the stellar parameters,
[Me/H], $T_{\rm eff}$,  and age ($\tau$). These are 
compared with those derived in other studies and in Sect. 2.3.3 we 
provide a detailed discussion of stellar ages determined from 
chromospheric indices. Sect. 3 presents and discusses 
selection effects in the age-metallicity
plot, the implications of our findings are 
 further discussed in Sect. 4. Finally Sect. 5 contains our conclusions.

\section{Data}

We selected all stars in the Hipparcos Catalogue (ESA, 1997) that had 
Str\"omgren $uvby$-photometry available in the large  catalogue compiled
by Hauck \& Mermilliod (1998). Further restrictions on which stars would
be allowed into our catalogue were imposed by the interval in  $b-y$,
$m_1$, and $c_1$
in which the calibrations of [Fe/H] and $T_{\rm eff}$ are valid, see 
Sect. \ref{metteff.sect}. Furthermore, we required all the stars to 
have a relative error in the parallax less than 25 $\%$. This was deemed
necessary in order to be able to derive reliable ages. The need to impose
this constraint will be further demonstrated in Sect. \ref{ages.sect}. 

\subsection{Binarity}

For binary stars we cannot determine stellar ages, neither
 effective temperatures nor metallicities
 by simply using the available stellar isochrones and available
calibrations of $uvby$-photometry. To ensure that the contamination of our 
sample from binaries is minimized we excluded all stars that were flagged 
either as proved binaries or as probable binaries in the Hipparcos Catalogue
(ESA 1997), e.g. all stars with CCDM number.
To be as conservative as possible we excluded all stars that were either
detected through the Hipparcos survey as binaries or had other indications of
binarity, e.g. a stochastic solution or were marked as suspected non-single.

Of the 14112 stars for which we have $uvby$-photometry 
 and for which ages could 
be determined, 3946 stars were rejected due to being binaries or potential
binaries. In this way we ensure that the stellar sample is as free as possible
from binaries. Obviously undetected close binaries can still be in the sample.
These can be discovered through spectroscopy. 

\subsection{Metallicity and effective temperature determinations}
\label{metteff.sect}

For each star we determine estimates of the interstellar reddening
toward the star by using the model of Hakkila et al. (1997). 
We also apply the Lutz-Kelker correction to the magnitudes using the
mean bias correction term
from Koen (1992) with index p=4 (see  Ng \&
Bertelli, 1998, for a discussion of the relevance of this correction
to age determinations). The  corrected magnitudes and colours are then
used to derive metallicities and effective temperatures. 

\paragraph{Metallicities} were derived using the calibration by 
Schuster \& Nissen (1989).
This calibration is valid for G dwarf stars for $0.37\leq(b-y)\leq0.59$ 
and for F dwarf stars for $0.22\leq(b-y)\leq 0.38$ (their Eq. (3) and
(2) respectively).

\paragraph{Effective temperatures} were derived using the calibration
in Eq. (9) in
Alonso et al. (1996). This calibration is valid for 
$ 0.25 \la b-y \la 0.7$.

By using these two calibrations we exclude a number of stars from further
study. The red cut in $b-y$ for the Schuster \& Nissen (1989) calibration
excludes a large number  of late K dwarf stars. This, however, is not a 
problem for us since it is not possible to derive ages for stars that 
far down on the main-sequence where all ages (for a given metallicity) are
degenerate. As can be seen from the limits given above 
 it is the determination of 
the metallicities
that imposes further cuts in our sample. 

\subsubsection{Comparison with other studies - Metallicities}

\begin{figure}
\resizebox{\hsize}{!}{\includegraphics{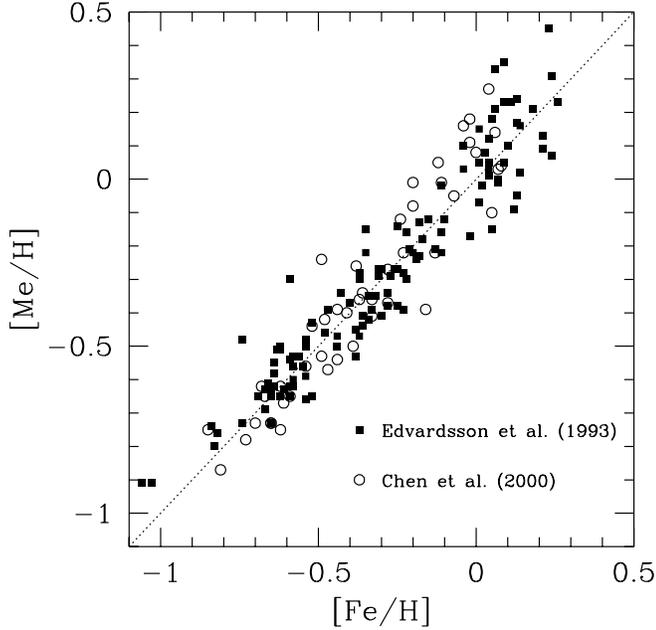}}
\caption[]{Comparison of photometric ([Me/H]) and spectroscopic 
([Fe/H]) determinations
of the stellar metallicity and iron abundance respectively with Edvardsson
 et al. (1993) and with Chen et al. (2000)}
\label{comp.ab.fig}
\end{figure}

We check the metallicities derived from the Str\"omgren photometry
(hereafter denoted [Me/H]) by comparing them with spectroscopic
measurements available in the literature (denoted [Fe/H]).  It is
especially important to make the comparison with large samples that
have been homogeneously treated. We have chosen two recent large
spectroscopic studies, Edvardsson et al. (1993) and Chen et al. (2000),
Fig. \ref{comp.ab.fig}.

The large catalogue by Cayrel de Strobel et al. (1997) was not
 considered here since this is a compilation of data from many
 different spectroscopic studies and although very valuable in many
 ways a comparison with that  catalogue would not  address the issue
 about the goodness of our metallicity determinations in a systematic
 way.

From the comparison we find that the systematic differences and scatters are 

 $$[{\rm Me/H}]_{\rm This~ work} - 
[{\rm Fe/H}]_{\rm E93}
 =+0.01\pm 0.10$$ and 

 $$[{\rm Me/H}]_{\rm This~ work} - 
[{\rm Fe/H}]_{\rm Chen00} = +0.02 \pm 0.11.$$ 

Schuster \& Nissen (1989) found that the uncertainty in their calibration 
of [Fe/H] was 0.16 dex. Our results, based on a comparison of spectroscopic
and photometric metallicities for a larger and more homogeneous 
sample then that used in Schuster \& Nissen (1998), in fact shows
 that their calibration is even
 better than 0.16 dex. We thus conclude that our [Me/H] are in very good
agreement with spectroscopically derived [Fe/H] and that we can expect the 
errors in derived ages from errors in the [Fe/H] determination to 
indeed be small.

\subsubsection{Comparison with other studies - Effective temperatures}

\begin{figure}
\resizebox{\hsize}{!}{\includegraphics{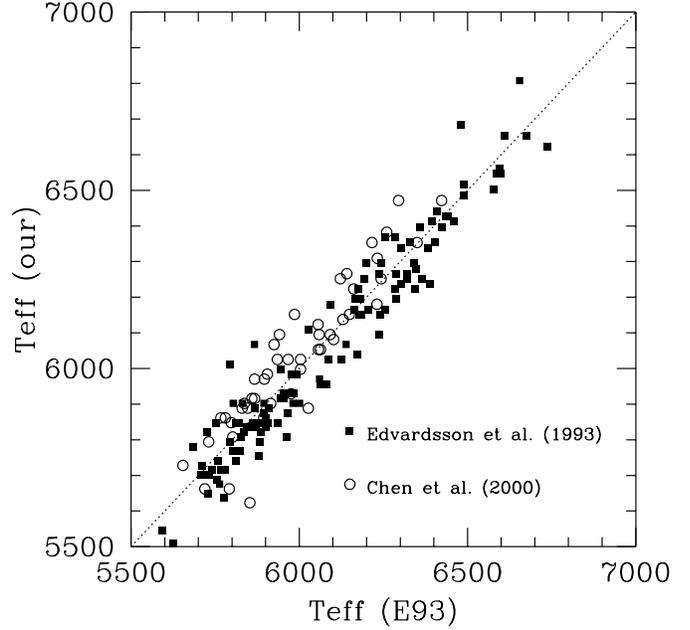}}
\caption[]{Comparison of effective temperatures 
derived in this work and in  Edvardsson et al. (1993) and with Chen et al.
(2000)}
\label{comp.teff.fig}
\end{figure}

In Fig. \ref{comp.teff.fig} we compare the effective temperatures we
derive using Str\"omgren photometry with those derived in Edvardsson
et al. (1993).  We find a mean difference of $-21$ K with a scatter of
72 K for the 120 stars in common between our full sample and that of
Edvardsson et al. (1993). Comparing with the 
Chen et al. (2000) sample we find a mean difference in effective
temperature of $+38$ K with a scatter of 80 K for the 
47 stars in common. We thus feel justified in considering our
effective temperatures well defined and that we can assume that
our effective temperatures, as  compared with spectroscopically
determined values, have an error of not  more than 100 K for the
majority of the stars.

\subsection{Determination of stellar ages}
\label{ages.sect}

To determine the age of each star in our sample we make use of the 
rapid stellar evolution algorithm presented by Hurley et al. (2000). 
This algorithm comprises a detailed set of evolution functions which 
cover all evolutionary phases from the zero-age main-sequence (ZAMS) up to, 
and including, remnant phases such as naked helium stars and white dwarfs. 
All the functions are analytic, except at a few special points, and
are valid for masses between 0.1 and $100 \, M_{\odot}$ and 
metallicities from $Z = 10^{-4}$ to 0.03. 
The functions themselves are derived from the grid of evolutionary
tracks produced by Pols et al.~(1998) using the detailed Eggleton
evolution code (Han et al. 1994).
These tracks include a modest amount of
enhanced mixing beyond the classical boundary of convective instability, i.e.
{\it convective overshoot}, and are approximated by the functions to 
within an accuracy of 5\% across the entire grid. 
The hydrogen and helium abundances are functions of the metallicity: 
$X = 0.76 - 3.0 Z$ and $Y = 0.24 + 2.0 Z$. 

For each star we have the observationally determined quantities of 
metal abundance, [Me/H], absolute magnitude, $M_{\rm V}$, and effective 
temperature, $\log T_{\rm eff}$ which define a point on the 
Hertzsprung-Russell diagram. 
The stellar evolution algorithm is then used to provide a theoretical mass 
and age appropriate to this point by minimizing the quantity 
\begin{equation}
\epsilon = \sqrt{\left( M_{{\rm V}, o} - M_{{\rm V}, t} \right)^2 
+ 16 \left( \log T_{{\rm eff}, o} - \log T_{{\rm eff}, t} \right)^2} \,  
\label{epsilon.eq}
\end{equation}
(cf. Ng \& Bertelli 1998). 
The subscripts $o$ and $t$ refer to observational and theoretical
quantities respectively. 
Assuming that the observational point does not lie below the ZAMS and can 
be reached within the maximum allowed evolution time, then for most 
stars $\epsilon$ is limited only by the tolerance chosen for the fitting 
process. 
This is because the evolution algorithm is continuous in mass, metallicity 
and age across the entire parameter space.  
We set the maximum allowed age to be $15.0\,$Gyr. 
We assume that all stars in our sample are either in the main-sequence (MS), 
Hertzsprung-gap or first giant branch phases of evolution, so in general 
the mass and age chosen is unique for the observational point. 
This is not always true if the point lies near the MS {\it hook} feature 
on an evolution track, as some overlap occurs in this region, but as this 
phase is rapid only a minimal error is introduced.  

The appropriate bolometric corrections required for the conversion of 
theoretical luminosity provided by the evolution functions to $M_{\rm V}$ 
are obtained from the synthetic stellar spectra computed by Kurucz (1992). 
We add to this an extra correction of --0.123 to account for the difference 
in solar offset between the models of Kurucz (1992) and Bessel et al. (1998). 
To convert from metal abundance to metallicity we use 
\begin{equation}
Z = 0.76 / \left( 3.0 + 37.425 \times 10^{\rm -[Me/H]} \right) 
\end{equation}
as given by Pols et al.~(1998). 
Mass loss is included in the stellar evolution algorithm but this 
option is not utilized here as the rates are uncertain and most stars 
in our sample are not evolved enough for mass loss to noticeably 
affect the derived ages. 

After fitting an age and a mass for each star in the standard data set we 
repeat the process taking the assumed errors in the data into account. 
These errors are $\pm 100\,$K in $T_{\rm eff}$  and 
$\pm 0.10$ in [Me/H]. For each star we also varied the $M_V$ according
to the $1 \sigma$  error in the parallax.
This gives 7 sets of age, mass and $\epsilon$ for each star from which an 
average age can be calculated according to the number of {\it good} fits. 
We arbitrarily determine a good fit to require $\epsilon \leq 0.02$, 
i.e. corresponding to a fitting error of 0.005 in $\log T_{\rm eff}$ if 
$M_{{\rm V}, o} = M_{{\rm V}, t}$.

\subsubsection{What is a good stellar age?}

What is a good age? Our method of determining the stellar ages allows 
several ways of determining this. The goodness of the fit is estimated
through Eq. (\ref{epsilon.eq}). As described in the previous section
we make seven age estimates for each star.
An age estimate is deemed good if $\epsilon < 0.02$ which corresponds 
to an error in $\log T_{\rm eff}$ of 0.005 if the input $M_V$ is the 
true $M_V$. 
The majority of the stars (8945 of  10166 stars or 88 \%) have six or seven 
good fits. 253 of the stars have no good fit. 
Stars with fewer good fits are not necessarily bad. For example,
a metal-rich very old star situated on the sub-giant branch might fall below
the stellar evolutionary tracks when the metallicity is decreased or the 
magnitude increased, thus yielding a bad fit. However, such a star is still
a ``good'' star and should be included into further considerations. 

\begin{table}
\caption[]{Stellar parameters derived in this work. n indicates the number of 
good fits.  The full table for 
all 5828 stars in our final sample is only available electronically }
\begin{tabular}{rlrrrrrrlrcr}
\hline\noalign{\smallskip}
HIP & M$_V$ & $\log T_{\rm eff}$ & [Me/H] & $\tau_{\rm mean}$  & $\sigma_{\tau}$ &  n &
 Mass \\
       &    &          & dex & Gyr & Gyr &  & M$_{\odot}$\\
\noalign{\smallskip}
\hline\noalign{\smallskip}
    20 & 3.48 & 3.801& --0.31 & 4.334 & 1.443 &  7 & 1.157 \\
    23 & 2.90 & 3.817& --0.36 & 2.565 & 0.478 &  7 & 1.321 \\
    33 & 2.64 & 3.818& --0.17 & 2.111 & 0.306 &  7 & 1.452\\ 
    39 & 2.49 & 3.817& --0.29 & 2.180 & 0.271 &  7 & 1.455\\ 
...& ...   &  ...   &  ...   & ...    & ... & ...    & ...    \\
...& ...   &  ...   &  ...   & ...    & ... & ...    & ...    \\
...& ...   &  ...   &  ...   & ...    & ... & ...    & ...    \\
\noalign{\smallskip}
\hline
\end{tabular}
\label{finalsample.tab}
\end{table}

\begin{figure}
\resizebox{\hsize}{!}{\includegraphics{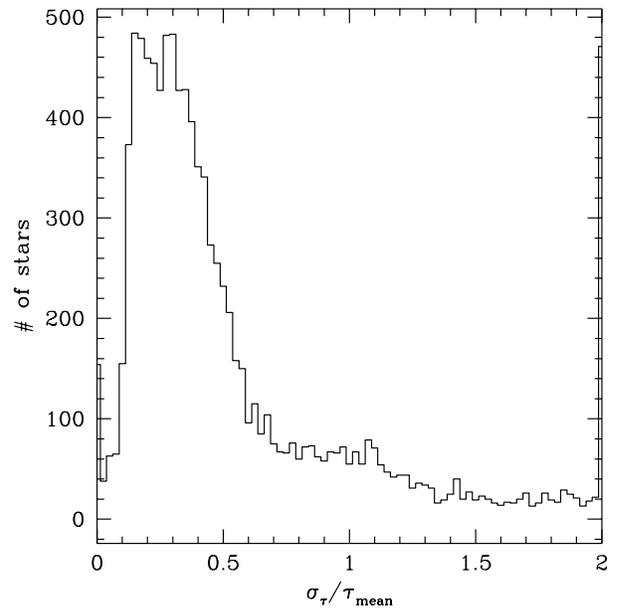}}
\caption[]{Histogram of $\sigma_{\tau}/\tau_{\rm mean}$ for all stars 
in the full sample, excluding the 253 stars that have no good fit. 
See the text for explanation of $\sigma_{\tau}$
and $\tau_{\rm mean}$}
\label{sigma.hist.fig}
\end{figure}

\begin{figure}
\resizebox{\hsize}{!}{\includegraphics{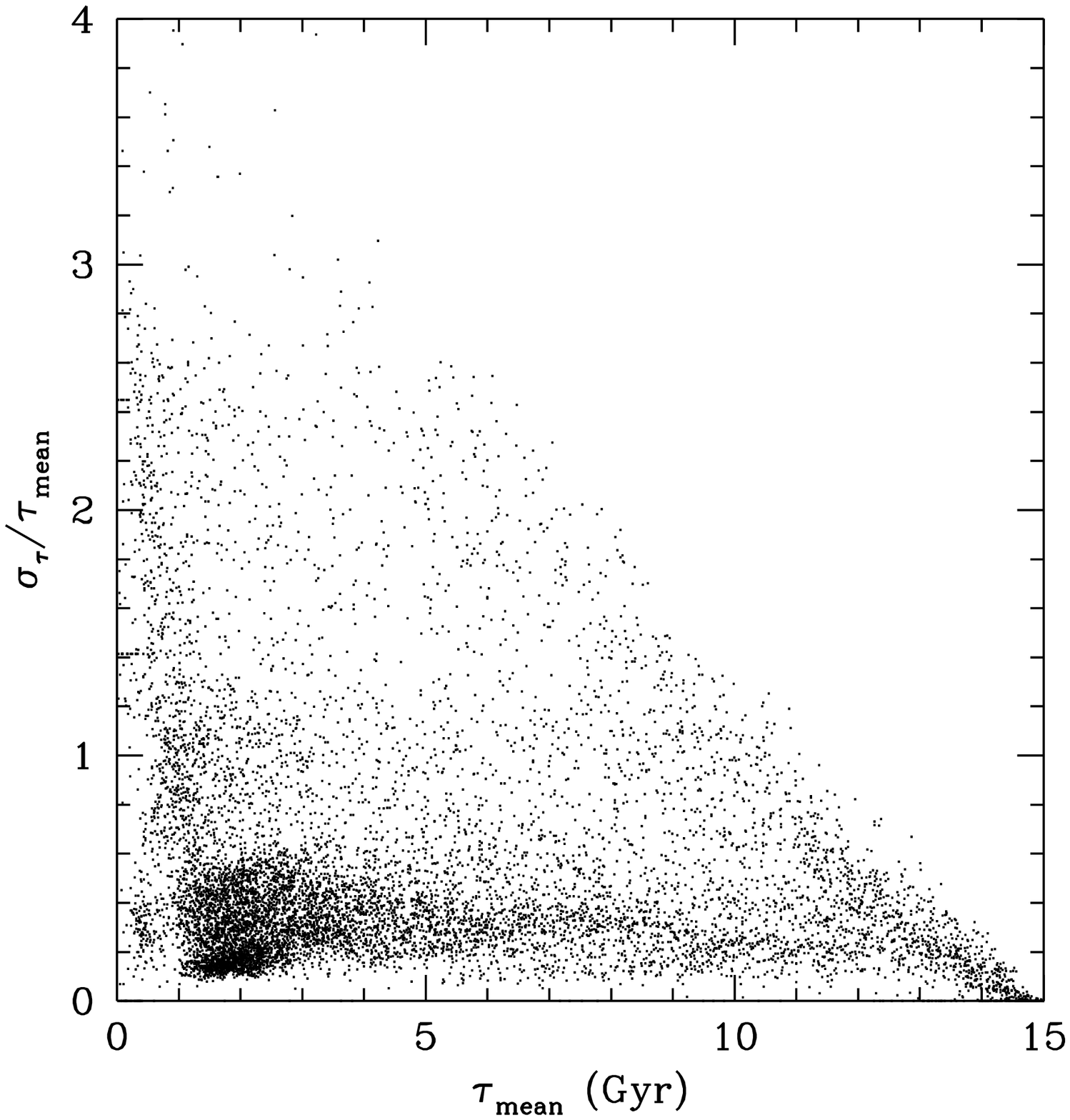}}
\caption[]{$\sigma_{\tau}/\tau_{\rm mean}$ as a function of 
$\tau_{\rm mean}$ for
all stars in the full sample}
\label{sigma.age.fig}
\end{figure}

\begin{figure}
\resizebox{\hsize}{!}{\includegraphics{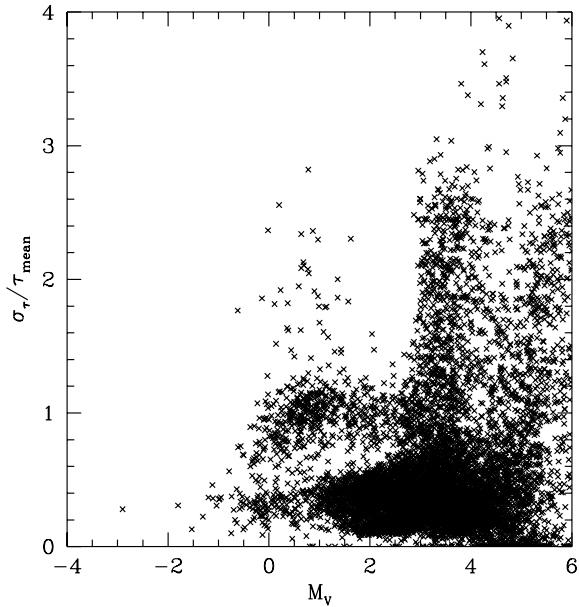}}
\caption[]{$\sigma_{\tau}/\tau_{\rm mean}$ as a function of 
absolute magnitude for
all stars in the full sample. }
\label{sigma.mv.fig}
\end{figure}

To quantify this we find the mean age for  each star using only the 
good fits. Then  we calculate the scatter around the mean
age,  $\sigma_{\tau}$. However, it is only the
$\sigma_{\tau}$ in relation to the mean age determined that give
useful information. Figure \ref{sigma.hist.fig} shows the  distribution
of $\sigma_{\tau}/\tau_{\rm mean}$ for all stars with more than 1 good
fit, i.e. when the concept of $\sigma_{\tau}$ becomes meaningful. 3744 stars
have  $\sigma_{\tau}/\tau_{\rm mean} > 0.5$. We will use 0.5 as the
dividing point between what we deem a good and a less good age. This
number is not magical in  any sense but inspection of individual stars
has lead us to the  conclusion that this is a reasonable
cut-off.
Figures \ref{sigma.age.fig} and \ref{sigma.mv.fig} show the variation
$\sigma_{\tau}/\tau_{\rm mean}$ as a function of mean age and absolute
magnitude. From now on we will only include stars with
$\sigma_{\tau}/\tau_{\rm mean} \leq 0.5$ and at least two good fits.

Determination of stellar ages becomes progressively more difficult
and uncertain as we proceed to fainter and fainter stars in the
HR-diagram because the stellar isochrones gets closer and closer.
For  too (absolutely) faint stars there will be a tendency to over-estimate
the stellar age.
To find out where our method finds its limit we have studied those 
Hyades stars present in our sample that are also in Perryman
et al. (1998). From this we find that our ages start to become 
systematically older below $M_V>4.4$. We will henceforth take this
as our lower limit for including stars into our discussions.

\subsubsection{Comparison with Ng \& Bertelli (1998)}
\label{ng.comp.sect}

\begin{figure}
\resizebox{\hsize}{!}{\includegraphics{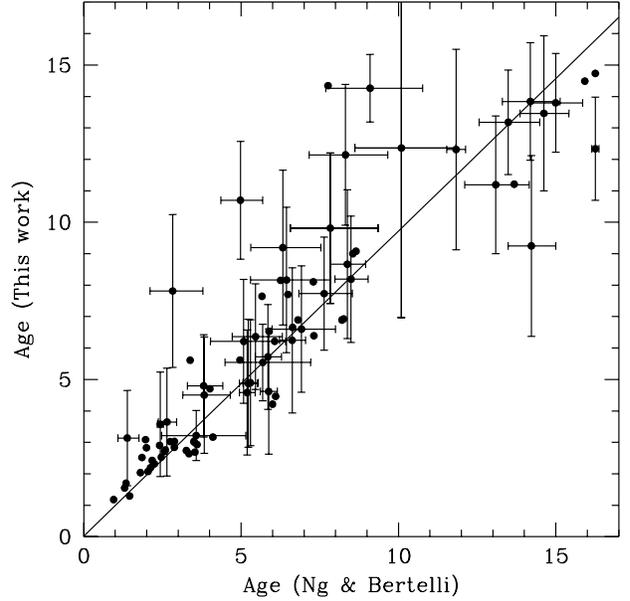}}
\caption[]{Comparison of ages 
derived in this work and in Ng \& Bertelli (1998). Errors bars are 
indicated if they are larger than 1.5 Gyr. Only stars 
that have $M_V<4.4$, $\sigma_{\tau}/\tau_{\rm mean}\leq 0.5$ 
and the number of good fits in both studies are larger than 1
are included. A linear least square fit, taking the indicated errors
into account, is also shown}
\label{comp.ages.fig}
\end{figure}

In Fig. \ref{comp.ages.fig} we compare the ages derived here with those
of Ng \& Bertelli (1998) for the Edvardsson et al. (1993) sample. This
is a particularly valuable comparison since we are basically using the
same method to determine our fits of the stellar tracks to the data
points  but use different sets of stellar evolutionary tracks, see
Pols et al. (1998) and Bertelli et al. (1994). The overall agreement is
good. We note that there are stars that deviate significantly
from the general trend. However, the age determinations for those 
stars also have large errorbars and most of them would, within the 
estimate error, fit well into the general trend. This is especially
true for the oldest stars. It is also to be expected that the oldest
stars should have the largest errors on their derived ages since
stellar iscohrones crowd relatively more for old ages
than for young.

\subsubsection{Isochrone vs chromospheric ages}
\label{comp.rochapinto.sect}

\begin{figure*}
{\resizebox{12cm}{!}{\includegraphics[angle=-90]
{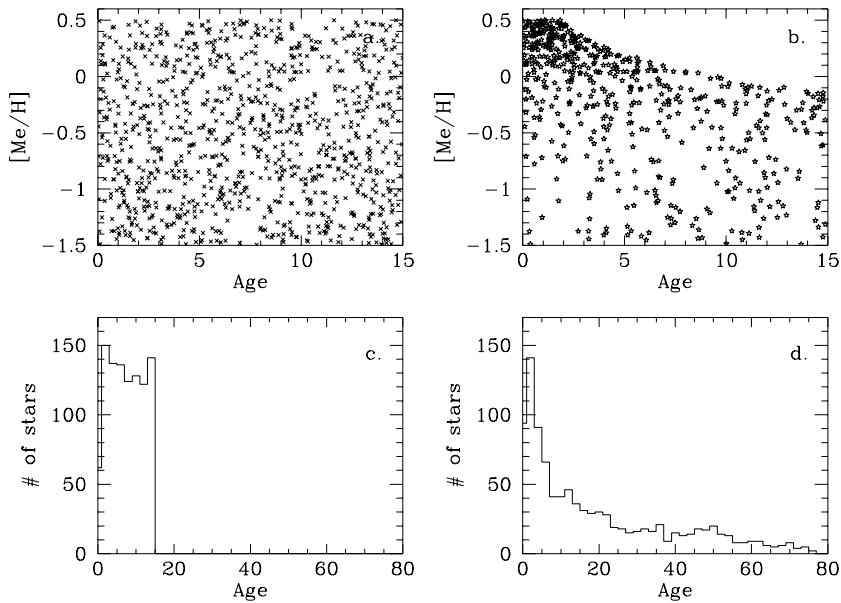}}}
\hfill
\parbox[t]{65mm}{
\caption[]{These panels show an illustration of the 
effects on the distribution of stellar ages when the calibration
from Rocha-Pinto \& Maciel (1998) is applied to a sample of stars.
{\bf a} A simulated age-metallicity plot (ages derived from chromospheric 
indices). 
{\bf b} the ages corrected according
to Eqns. (3) and (4) in Rocha-Pinto \& Maciel (1998). 
{\bf c} and {\bf d} show 
the respective histograms of the ages for plot {\bf a} and {\bf b}}
\label{fake.amr.fig}}
\end{figure*}

Stellar evolutionary tracks and isochrones are not the only way
to determine stellar ages. Especially the method of using 
chromospheric indices to date stars has recently been applied
in studies of the star-formation history and age-metallicity
relation of the stellar disk in the Galaxy.
The results have been presented in a series of papers by
Rocha-Pinto and co-authours, e.g. Rocha-Pinto et al. (2000).
 In the context of the stellar
ages the paper by Rocha-Pinto \& Maciel (1998) is the important
paper which details the exact way the ages are determined.

In brief their method relies on one of several possible relations
derived by Soderblom et al. (1991)  between the $R'_{\rm HK}$ index
and the age of a star, $\tau_{\rm chr}$.  This relation is, in
Rocha-Pinto \& Maciel (1998), calibrated onto isochrone ages,
$\tau_{\rm iso}$.  The calibration is done using 40 stars in common
between their sample and that of Edvardsson et al. (1993). The more
up-to-date ages from Ng \& Bertelli (1998) for the Edvardsson et
al. (1993) sample were used.  The difference of the $\log \tau_{\rm
chr}$ and $\log \tau_{\rm iso}$ is found  as a function of [Fe/H]
(Fig. 1b in Rocha-Pinto \& Maciel 1998). This  correction is then
applied to the chromospheric ages for their stars. There are three
points worth noting about this calibration: 1) the relation found by
Soderblom et al. (1991) is by no means unique (see discussions in
Soderblom et al.), 2) the calibration between the difference of $\log
\tau_{\rm chr}$ and $\log \tau_{iso}$ as a function of [Fe/H] is not
well defined, another polynomial (even a  straight line) could fit
equally well (no errors on the fitting are given), 3)  it is not clear
that the fitting formula does not suffer from selection  biases,
i.e. a small number of stars are involved and the scatter is large.
With regards to the last point we note in particular the the
correction formula found in Rocha-Pinto \& Maciel (1998), because it
is the  $\Delta \log \tau$ that is determined, applies a correction
{\sl factor}  to the ages such that a large age will get a large
correction factor  while a small age, with the same [Me/H], will only
have a small correction.  Moreover, the correction formula is so
constructed that no matter where the  star is originally in the
age-metallicity plot, old, metal-rich stars will have a large
correction factor, making them younger, and young, metal-poor stars
will have a small correction factor, i.e. making them older. In this
way all metal-rich old stars will automatically become  young and the majority of 
young, metal-poor will become old. 
Thus a one-to-one  age metallicity
relation will automatically be found using the derived correction
formula no matter how the stars are first distributed in the
age-metallicity diagram.

As a simple illustration of the described effect using the corrections
 according to Eqns. (3) and (4) in Rocha-Pinto \& Maciel (1998) we
 have randomly distributed stars in an age-metallicity plot and then
 applied the correction factor to these (chromospheric) ages. The
 results are exemplified in  Fig. \ref{fake.amr.fig}. As can be seen
 here the correction factor moves  all stars in the upper right hand
 corner to the left and the majority of the  stars in the lower left
 hand corner are moved to the right. Many are moved  altogether
 outside the plot and in Rocha-Pinto \& Maciel (1998) these stars are
 disregarded from further study.

\begin{figure}
\resizebox{\hsize}{!}{\includegraphics{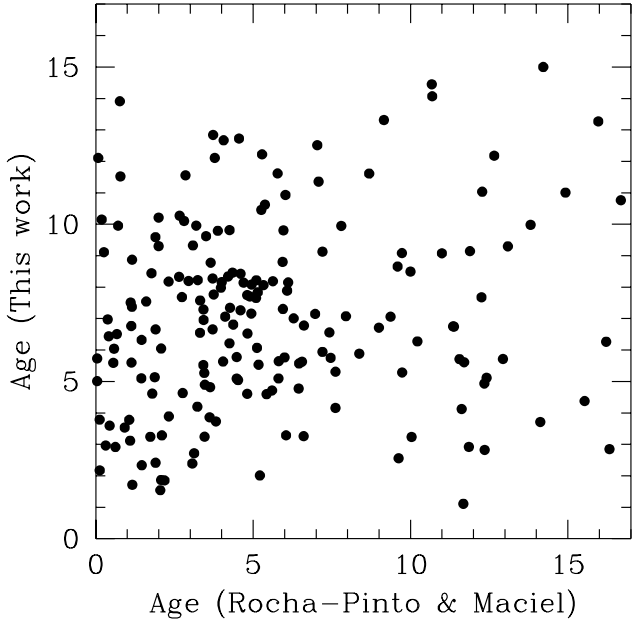}}
\caption[]{Comparison of ages 
derived in this work and in Rocha-Pinto \& Maciel (1998). Only stars
with more than one good fit and with $M_V<4.4$ and 
$\sigma_{\tau}/\tau \leq 0.5$ are included
}
\label{comp.ages.rp.fig}
\end{figure}

We  have made extensive comparisons of the ages we derive from
stellar evolutionary tracks with those determined by Rocha-Pinto \&
Maciel (1998).  In total there are 261 stars in common between their
work and that presented here. For  stars with  $M_V<4.4$ and 
$\sigma_{\tau}/\tau \leq 0.5$ (in our determination) we compare 
the chromospheric ages with our ages derived from stellar 
evolutionary tracks in Fig. \ref{comp.ages.rp.fig}.

We were especially concerned to check that our method did not give
spurious old ages when in reality the star was young, as indicated
by the chromospheric activity. To check this we simply plotted the stars
in the HR-diagram together with stellar evolutionary tracks from 
Bertelli et al. (1994).

\begin{figure}
\resizebox{\hsize}{!}{\includegraphics{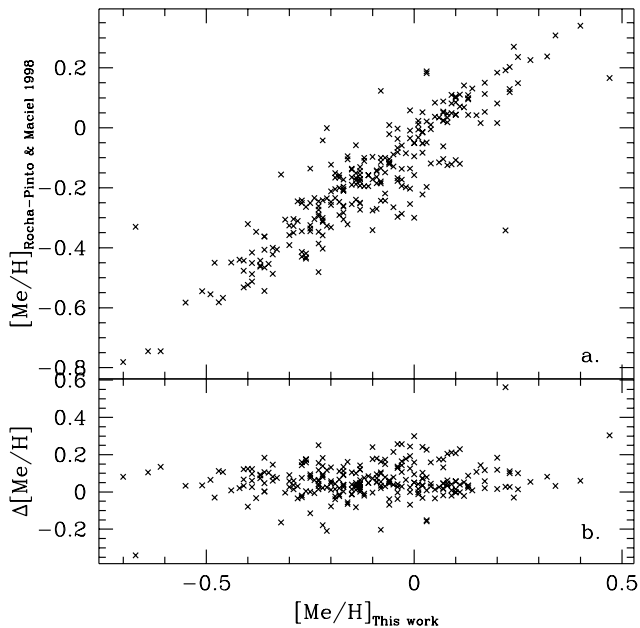}}
\caption[]{Comparison of metallicities derived by us and by
Rocha-Pinto \& Maciel (1998). In panel b. we show the difference
between the two determinations in the sense {\sl our results}-{\sl
their results}}
\label{comp.meh.rp.fig}
\end{figure}

 For completeness we also compare the [Me/H] derived by us
and by Rocha-Pinto \& Maciel (1998), Fig. \ref{comp.meh.rp.fig}. The
agreement is good. The difference between the two studies is $0.06
\pm 0.09$. This is before applying
corrections according to Eqn. (5) in Rocha-Pinto \& Maciel
(1998). However, this correction is only for chromospherically active
stars (compare their Figure
5). Most of the stars in common between their study and our  are
almost all inactive stars, so this correction would have little
influence on the overall comparison of our results.

We want in particular 
to stress here that we do {\bf not} say that stellar ages can not be
derived reliably from chromospheric indices. Especially for young
stars this method appears very attractive and would be a much needed
complement to isochrone ages since it is notoriously difficult, not to
say impossible, to derive reliable isochrone ages below $\sim 1 $ Gyr 
due to the crowding of isochrones in the HR-diagram. However, we are 
not equally convinced that they can be used to derive stellar ages
for the oldest stars in the galactic disk and would, in fact, caution
against this.

\section{The age-metallicity plot}

\begin{figure*}
{\resizebox{12cm}{!}{\includegraphics[angle=-90]{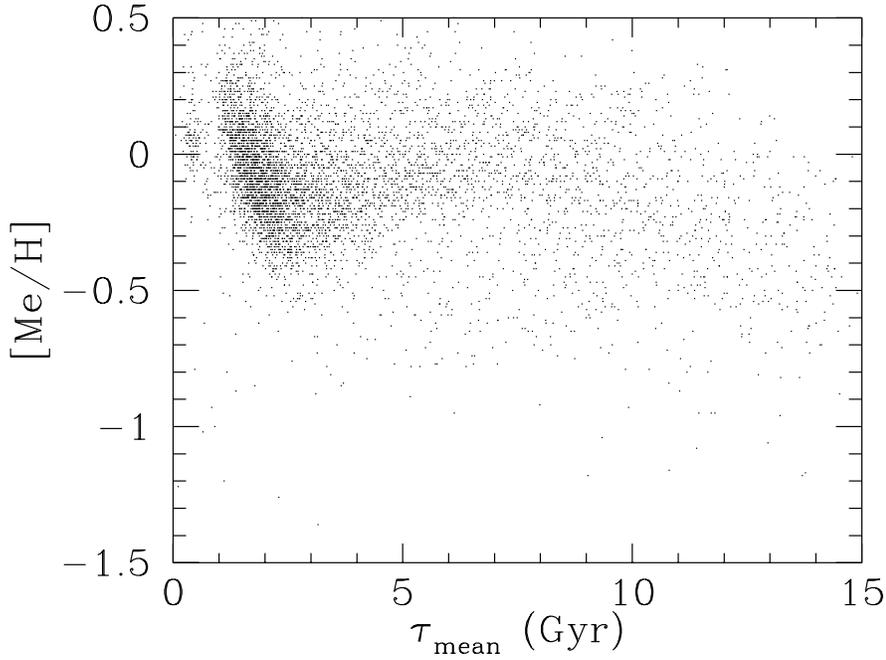}}
\hfill
\parbox[t]{65mm}{
\caption[]{The age-metallicity plot of all stars with $\sigma_{\tau}/\tau 
\leq 0.5$ and $M_V<4.4$. 5828 stars were included.
}
\label{amr_sel.fig}}}
\end{figure*}

We now turn to consider the age-metallicity plot of local disk stars.
Figure \ref{amr_sel.fig} shows the plot including all stars in our sample with 
errors in the mean less than 50 \% and $M_V<4.4$.
The plot includes 5828 stars.

The main impressions from this plot are the large spread both in age and 
metallicity as well as a pronounced structure around 2 Gyr. Is what we are 
seeing the true picture or are we here subject to influences of different
selection effects?

\subsection{The old, metal-rich stars}
\label{oldmet.sect}

\begin{figure*}
{\resizebox{12cm}{!}{\includegraphics[angle=-90]
{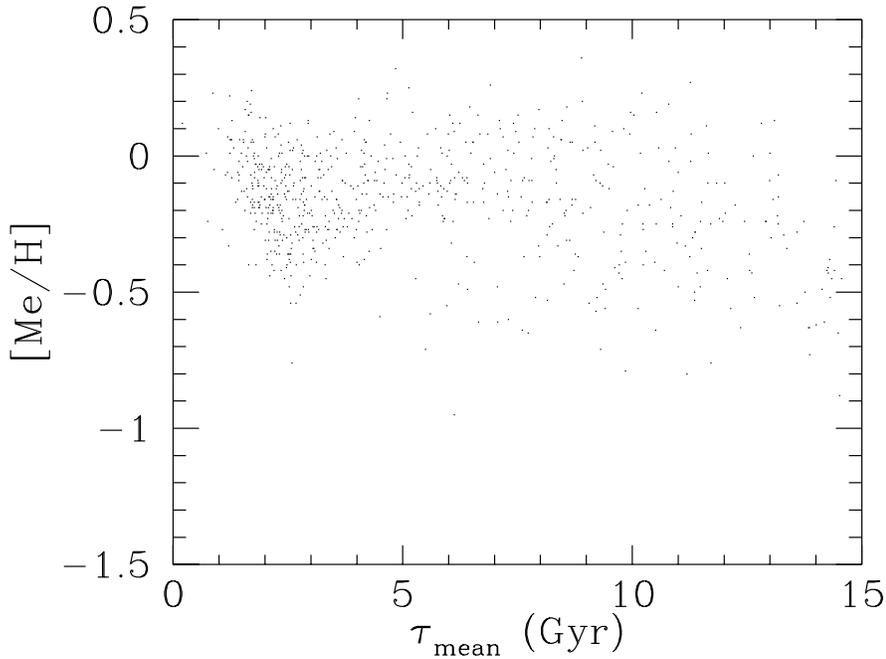}}
\hfill
\parbox[t]{65mm}{
\caption[]{The age-metallicity diagram for all stars with 
$\sigma_{\tau}/\tau 
\leq 0.5$, $M_V<4.4$, and $E(b-y) <0.005$. 716 stars were included.
 }
\label{amr_sel_eby.fig}}}
\end{figure*}

One of the most prominent features of  Fig. \ref{amr_sel.fig} is the
very real presence of stars that are both metal-rich, i.e. [Me/H]$>
0.0$, and old, i.e. with an age greater than 10 Gyr. The presence of
such stars, if real, clearly indicates that the star formation history
of the disk has been very complicated with regions (perhaps in the
inner disk?) that quickly reached a high metallicity.

Some of our high age objects are at the bottom of the giant branch in
the HR-diagram and would thus have somewhat more uncertain ages,
even though their formal errors are small, due to the difficulties in
the exact modeling of this region.
However, all the stars with high metallicities and high ages are turn-off
objects and are thus not susceptible to such errors in the age determination.

We have applied reddening corrections to the observed magnitudes for 
all our stars before deriving their stellar parameters. Our reddening
estimated are based on the model by Hakkila et al. (1997). In order
to check if this model has in any way systematically over-corrected the
reddening for stars that are metal-rich we select all stars with 
$E(b-y) < 0.005$, i.e. virtually negligible corrections, and plot the 
age-metallicity
diagram for these 716 stars in Fig. \ref{amr_sel_eby.fig}. Although much
fewer stars now appear in the plot as compared to Fig. \ref{amr_sel.fig}
the prominent features remain. In particular we note that old and metal-rich
stars remain in roughly equal numbers proving the reality of such stars.

\subsection{The F, G and K dwarf stars}

\begin{figure}
\resizebox{\hsize}{!}{\includegraphics{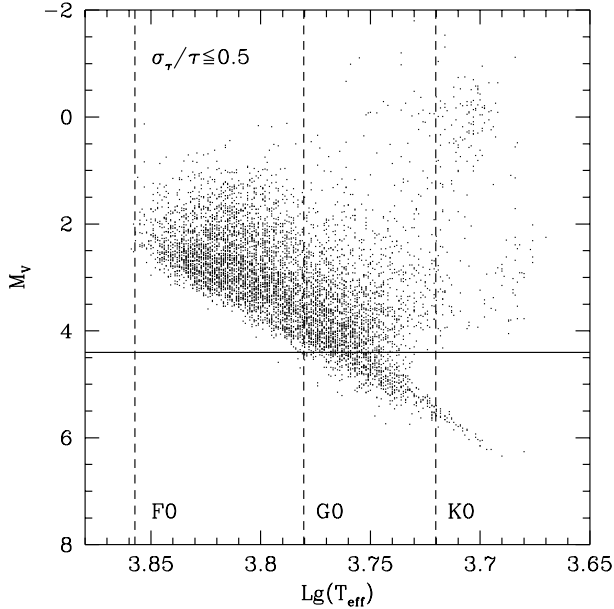}}
\caption[]{The HR-diagram of all stars with $\sigma_{\tau}/\tau \leq 0.5$
 and n$\geq 2$.
The major spectral class divisions are also indicated (taken from Binney \&
Merrifield 1998).}
\label{hr.fig}
\end{figure}

Figure \ref{hr.fig} shows the HR-diagram for all the stars in
Fig. \ref{amr_sel.fig}.  The major spectral classes are also
indicated. Although the majority of the stars  are turn-off stars we
also still include a few stars that are tending towards the giant
branch (i.e. $ 4.4<M_V<2 $ and $\lg T_{\rm eff}<3.7$) as 
well as a number of stars at
 $ 2<M_V $ and $\lg T_{\rm eff}<3.75$. We note that 
the latter stars all have very young ages, i.e. these are the stars that
make up the small concentration of stars at $\tau_{\rm mean} \sim 0.5$ and around
[Me/H] = 0.05 dex in Fig. \ref{amr_sel.fig}. When strong constraints 
on the reddening is imposed these stars are virtually all excluded.
 We let these stars remain
in the sample for now. The stars that are close to the bottom of 
the giant branch all have ages larger than 10 Gyr and scatter around 
 $-0.1$ dex. Some of these will remain also when stronger constraints
are set on the reddening. We keep also these stars in the full sample.
There are also a few stars that appear
``below'' the main-sequence. These are the most metal-poor stars in
our sample.

\begin{figure}
\resizebox{\hsize}{!}{\includegraphics{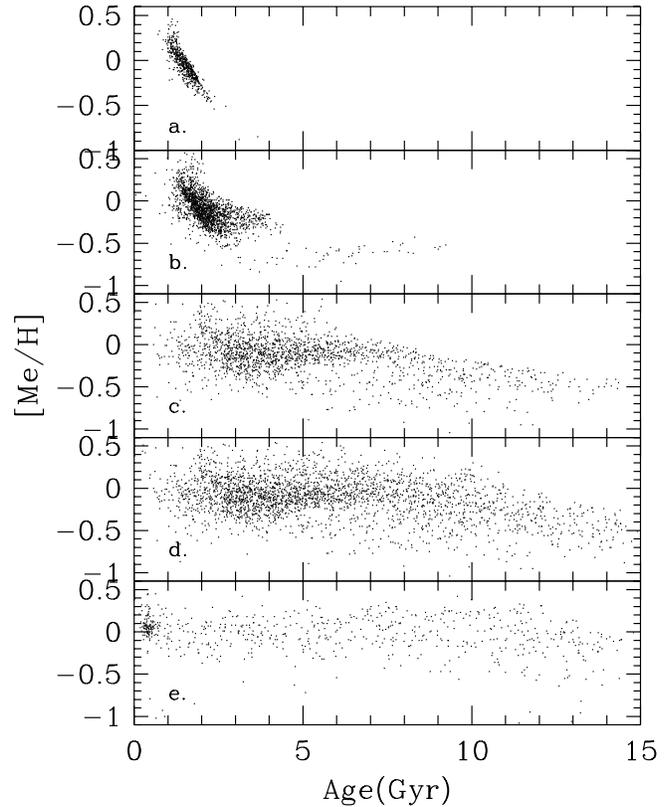}}
\caption[]{Age-metallicity plots of different sub-samples from Fig. 
\ref{amr_sel.fig} and Fig. \ref{hr.fig}.
{\bf a} stars with $\log T_{\rm eff}  > 3.83$, 
{\bf b} stars with $3.83 \geq \log T_{\rm eff}  > 3.8$, 
{\bf c} stars with $3.8 \geq \log T_{\rm eff}  > 3.77$, 
{\bf d} stars with $3.8 \geq \log T_{\rm eff}  > 3.75$, 
{\bf e} stars with $3.75 \geq \log T_{\rm eff} $. 
For clarity we only show [Me/H]$ > -1$. All stars have,
as before, $M_V<4.4$ and $\sigma_{\tau}/\tau_{\rm mean} \leq 0.5$
 and n$\geq 2$}
\label{amr_fgk.fig}
\end{figure}

Different types of selection effects are amply illustrated in
Fig. \ref{amr_fgk.fig} where we compare the age-metallicity plots of
several subsamples of  Fig. \ref{amr_sel.fig}. The stars are divided
into sub-groups through their effective temperature. The sub-groups
have been chosen as to emphasize the important differences. Figure
\ref{amr_fgk.fig} {\bf a} shows the stars with the highest effective
temperatures, $\log T_{\rm eff} > 3.83$. These  are all early F
stars. Their age-metallicity plot shows a narrow feature with age and
metallicity well correlated. Figure \ref{amr_fgk.fig} {\bf b} shows
stars in a lower temperature range, $3.83 \geq \log T_{\rm eff}  >
3.8$.  There is still a strong correlation between age and
metallicity. If this was the stellar sample we had selected for a
study of the age-metallicity relation we would derive a rather steep
and strong correlation between age and metallicity. This is basically
a late  F star sample.

Figure \ref{amr_fgk.fig} {\bf c} and {\bf d} show perhaps the most
interesting selection effects. Figure \ref{amr_fgk.fig} {\bf c} is
essentially a sample  centered around spectral class G0, $3.8 \geq \log
T_{\rm eff}  > 3.77$, while  Fig. \ref{amr_fgk.fig} {\bf d} shows the
same stars but also adds cooler G dwarf stars, $3.8 \geq \log T_{\rm
eff}  > 3.75$. If the first sample had been our chosen stars we would
derive a fairly strong one-to-one relation between age and
metallicity.  If the larger sample was added, the scatter would have
been increased, although the  error in the mean would be lower as more
stars were added, and a somewhat  shallower age-metallicity relation
derived.

Finally, Fig. \ref{amr_fgk.fig} {\bf e} shows all stars cooler than
$\log T_{\rm eff}=3.75$.  This includes not only dwarfs, compare the
HR-diagram, but also the brightest evolved  stars discussed in the
first paragraph of this section. These stars all have young ages and
are tracers of a stellar population, i.e. stars earlier than $\sim$
F0, that is excluded from this study due to the limitations of the
calibration of $T_{\rm eff}$.  These stars appear in the
age-metallicity plot as the small concentration of stars with around
solar metallicities and have the lowest ages of all.

The most important conclusion that we reach from studying
Fig. \ref{amr_fgk.fig} is that as we allow cooler and cooler stars to
enter our sample the upper right-hand corner of the age-metallicity
plot is progressively filled in. This  can be understood by
considering the stellar isochrones in
e.g. Fig. \ref{iso.fig}. As is illustrated there stellar
isochrones of more metal-rich stars have a redder as well as fainter
turn-off for a  given age than more metal-poor stars. This means that
metal-rich, old stars are excluded from the sample if we for example
set the cut in $\log T_{\rm eff}$ at $\sim$3.75. The  exact limits
vary with metallicity. Also remember that the stellar  sample we are
discussing here only includes stars that have $M_V < 4.4$.

\begin{figure}
\resizebox{\hsize}{!}{\includegraphics{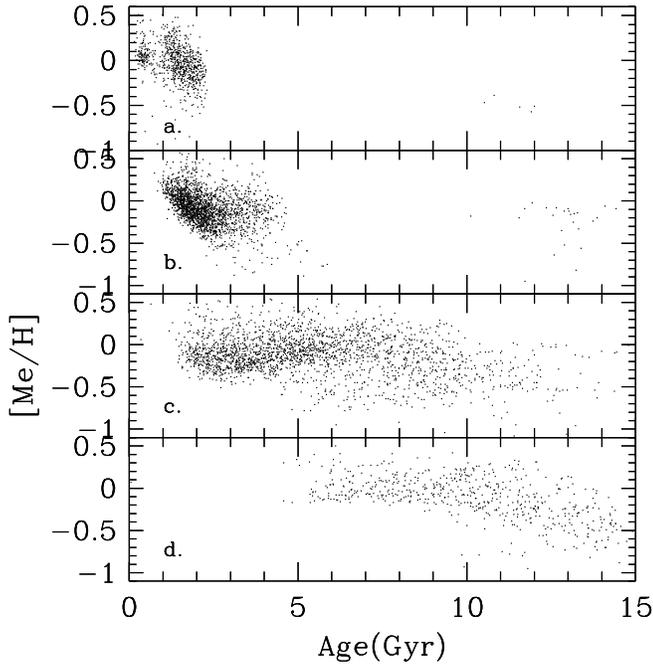}}
\caption[]{Age-metallicity plots of different sub-samples from 
Fig. \ref{amr_sel.fig} and Fig. \ref{hr.fig}.
{\bf a} stars with $M_V<2$, {\bf b} stars with $2\leq M_V <3$,
{\bf c} stars with $3 \leq M_V < 4$, {\bf d} stars with $4 \leq M_V < 4.4$. 
All stars have,
as before $\sigma_{\tau}/\tau_{\rm mean} \leq 0.5$ and n$\geq 2$}
\label{amr_MV.fig}
\end{figure}

\begin{figure}
\resizebox{6.2cm}{!}{\includegraphics{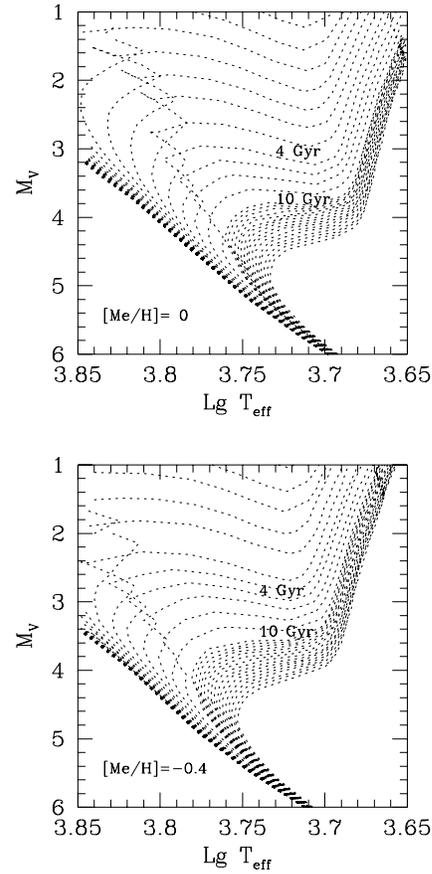}}
\caption[]{Theoretical isochrones from Bertelli et al. (1994) illustrating 
the different turn-off colours for metallicities as indicated
as well as the regions where stellar isochrones become too closely
packed as to prohibit determinations of stellar ages}
\label{iso.fig}
\end{figure}

Another illustration of selection effects is shown in
Fig. \ref{amr_MV.fig}.  In this figure we plot the ages and
metallicities for different bins of $M_V$.  A similar pattern as in
Fig. \ref{amr_fgk.fig} appears, but now the progress from a young,
confined population to a spread out is found  by going to fainter and
fainter samples. The selection effect is perhaps  most striking in the
last sub-plot where stars with $4.4 <M_V< 4$ are  shown. This
apparently  strange selection effect is purely due to the fact that it
becomes progressively harder to derive the stellar age as we move down in
magnitude. Thus these stars are simply excluded through our criteria
for  a good stellar age. That the effect stops at younger ages for
metal-rich  stars than for metal-poor once again can be understood from
considering the  stellar isochrones in
e.g. Fig. \ref{iso.fig}. From this  figure we see that the
isochrones in the band of absolute magnitudes between 4 and 4.4 are
overall more closely packed for the metal-poor isochrones (i.e. the
lower panel) than in the panel showing stellar isochrones with  solar
metallicities. This means that the stellar ages derived from  fitting
of evolutionary tracks necessarily will be more uncertain, and indeed
as illustrated in Fig. \ref{amr_MV.fig} no stars in our sample which
are in this faintest bins and have an age smaller than $\sim$ 10 Gyr
get an age estimate that has $\sigma_{\tau}/\tau \leq 0.5$.

From these investigations we conclude that the exact definition of the
stellar sample is of the highest importance.  It thus appears unlikely
that a stellar sample of the solar neighbourhood can be defined such
that the true age-metallicity plane is well sampled.

In studies of the age-metallicity plot and in the  subsequent
derivation of an age-metallicity  relation the volume completeness of
the sample is of prime importance.  For our sample no selection
according to distance was made. The important question is thus how
this influences the age-metallicity plot that we  have presented
e.g. in Figs. \ref{amr_sel.fig} and \ref{amr_sel_eby.fig}. 

 For our final
sample it is virtually impossible to assess the completeness due to
the complex route taken to arrive at the final sample. That is we only
include stars with $uvby$ which also have a relative error
in the parallax less than 0.25. Finally we impose a cut in the relative
errors of the derived ages.

\begin{figure}
\resizebox{6.2cm}{!}{\includegraphics{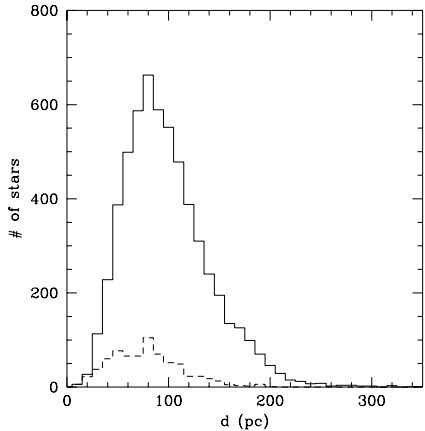}}
\caption[]{Histogram of the distances for stars in our final sample
(full line) and for the sub-sample of stars that have $E(b-y)<0.005$}
\label{hist_dist.fig}
\end{figure}

Thus, instead of trying to define a
subsample that is volume complete we will select stars 
with  $E(b-y)<0.005$ and  divide that sample
according to the distance. This sample has a minimum of
error sources for their age and metallicity determinations (see
previous investigation of effect from $E(b-y)$). 
 The distributions of distances for our full
 sample and  that for stars with negligible reddening are shown in
Fig. \ref{hist_dist.fig}.

\begin{figure}
\resizebox{\hsize}{!}{\includegraphics{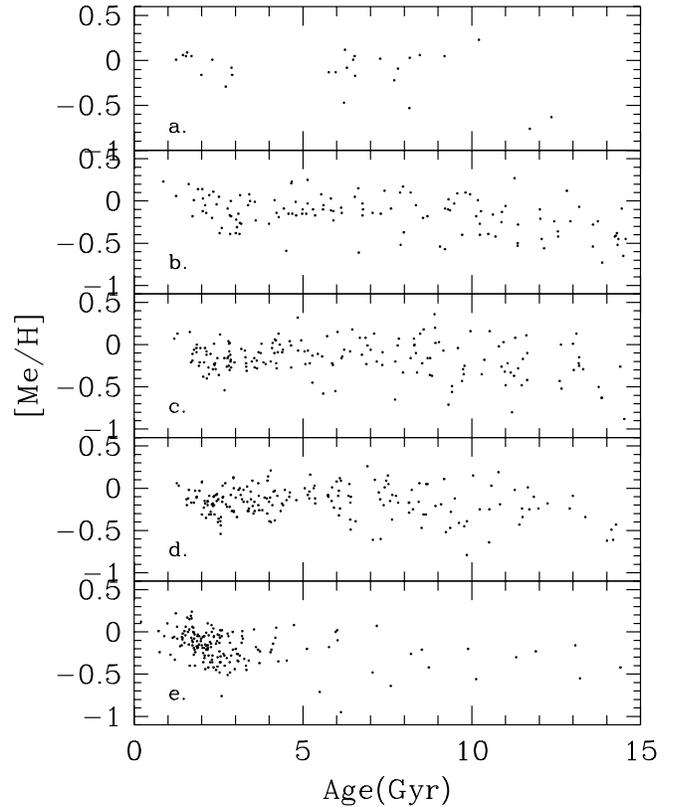}}
\caption[]{Age-metallicity plots of  sub-samples in different distance
ranges. 
{\bf a} stars with  $d<25$ pc, 
{\bf b} stars with  $25\leq d<50$ pc,
{\bf c} stars with  $50\leq d<75$ pc,
{\bf d} stars with  $75\leq d<100$ pc,
{\bf e} stars with  $100\leq d$ pc,
 All stars have,
as before, $M_V<4.4$ and $\sigma_{\tau}/\tau_{\rm mean} \leq 0.5$
 and n$\geq 2$ as well as $E(b-y)<0.005$}
\label{amr_dist.fig}
\end{figure}

The age distributions we find for different distance intervals in our
sample are essentially flat, Fig. \ref{amr_dist.fig}.  All 
distance bins  show large and roughly  similar
scatter at all, and especially old, ages.  
We note that the relative number of young stars increases with distance. 
This is due to the fact that they are intrinsically bright and we are thus
able to sample them better than the older, fainter stars. We also note
that in the last distance bin the old stars have a somewhat lower mean
metallicity than in the other bins. This is due to the fact that metal-poor stars,
for the same age, are intrinsically brighter than the metal-rich ones and
can thus be sampled more completely at larger distances. 

\begin{figure}
\resizebox{\hsize}{!}{\includegraphics{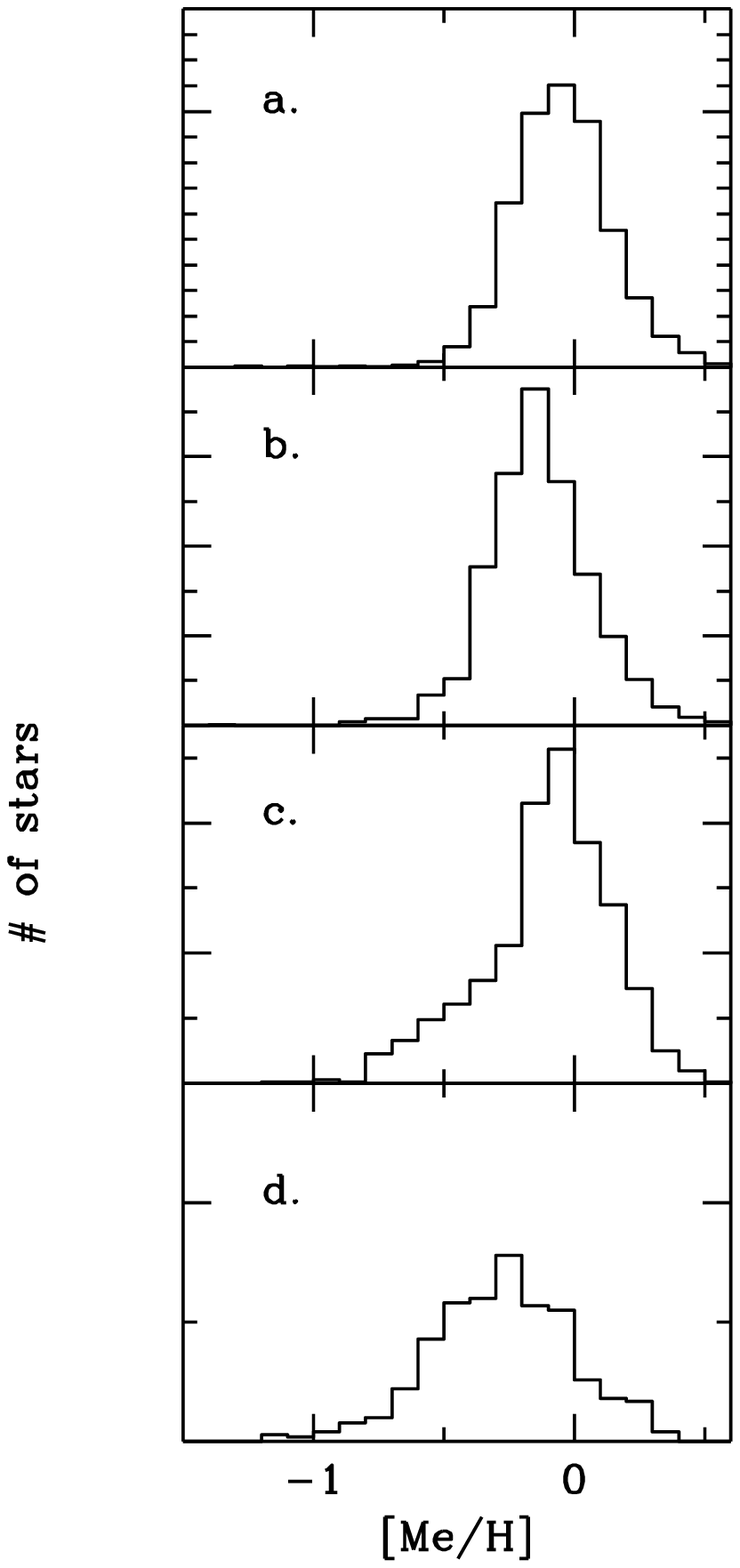}}
\caption[]{Histograms for four age-ranges, {\bf a} stars with ages
less than 2.5 Gyr, {\bf b} ages between 2.5 and 5 Gyr, {\bf c} ages between 
5 and 10 Gyr, and finally {\bf d} ages larger than 10 Gyr. Note 
the different ranges on the y-axes, each tick mark, in each
separate panel, corresponds to 
a unit of 50 stars. All stars have,
as before, $M_V<4.4$ and $\sigma_{\tau}/\tau_{\rm mean} \leq 0.5$ and n$\geq 2$. }
\label{histograms.fig}
\end{figure}

To complete the characterization of our sample we also show the distributions
of [Fe/H] for four different age ranges in Fig. \ref{histograms.fig}. 
The first bin contains all stars younger than 2.5 Gyr and has $<$[Me/H]$> =
-0.044$ with $\sigma$ equal to 0.19 dex. The second bin contains all
stars between 2.5 and 5 Gyr old and has  $<$[Me/H]$> =
-0.13$ with $\sigma$ equal to 0.20 dex. Third bin shows the
stars between 5 and 10 Gyr and has $<$[Me/H]$> =
-0.11$ with $\sigma$ equal to 0.26 dex, while the final bin contains
all stars older than 10 Gyr and has $<$[Me/H]$> =
-0.31$ and a $\sigma$ of 0.33 dex. All bins only contain stars with 
$M_V<4.4$ and $\sigma_{\tau}/\tau_{\rm mean} \leq 0.5$.

From this we may conclude that indeed the mean metallicity in the local
stellar disk is high for all ages and that the scatter in metallicity
at all ages also is high. It is also worth noting that the scatter in
metallicity at the highest ages cannot be due to an over-estimate of 
the reddening correction, see Sect. \ref{oldmet.sect} and Fig. 
\ref{amr_sel_eby.fig}, and hence we may conclude that the scatter we
see is a viable representation of the underlying true metallicity
distribution and that, most importantly, the scatter is {\sl at least}
as large as seen in diagram d of Fig. \ref{histograms.fig}. 

\section{Discussion -- Does a unique age-metallicity relation exist? }

The previous sections presented the resulting age-metallicity plot 
for the solar neighbourhood using 5828
 stars with well determined 
ages and metallicities. We also showed how different selection criteria
may bias the final conclusions. We will now proceed with a discussion of 
previous major determinations of the age-metallicity diagram available 
in the literature as well as to put our results into the context
of the galactic chemical evolution.


The most influential studies of the age-metallicity relation  are
perhaps Twarog (1980b) and Edvardsson et al. (1993).  The selection
of stellar samples as well as  the resulting conclusions in the two
papers are quite different and merits a closer look.

\begin{figure}
\resizebox{\hsize}{!}{\includegraphics{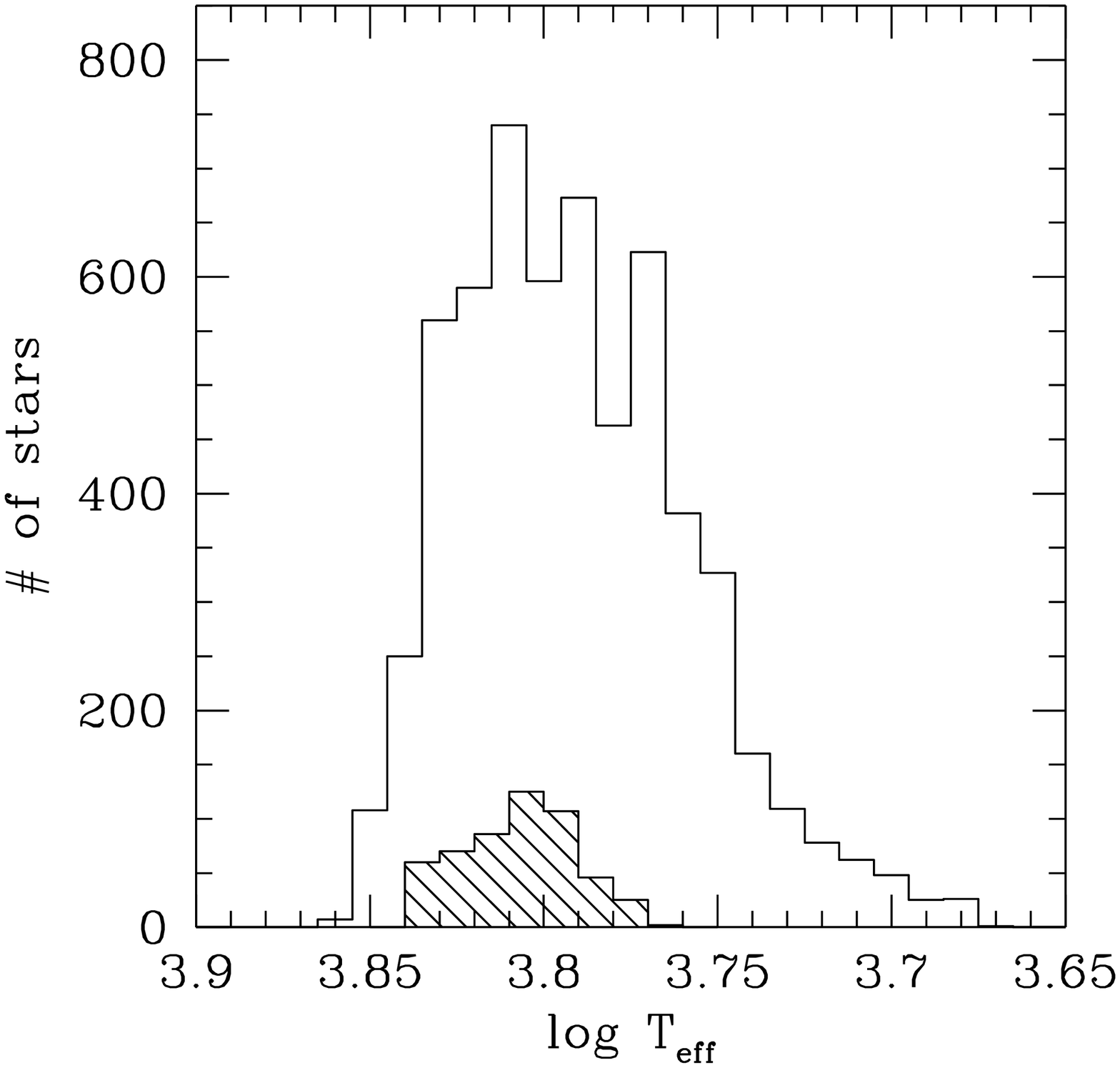}}
\caption[]{Histogram showing the distribution of effective temperatures
for the stars in Fig. \ref{amr_sel.fig}. The small, shaded histogram
shows the distribution for the sample in Meusinger et al. (1991) (reproduced
after their Fig. 1b).}
\label{teff_hist.fig}
\end{figure}

Twarog (1980a, 1980b) selected a stellar sample of F dwarf stars.  Twarog
(1980b) used a combined sample of carefully selected stars from both
the Michigan catalogue as well as using the  sample by Gr{\o}nbech \&
Olsen (1976, 1977).  The stellar $uvby-\beta$ photometry was later
reanalyzed in detail by Meusinger et al. (1991) using a revised
calibration of stellar metallicity and new stellar isochrones from
VandenBerg (1985).  The total, combined sample included 536 stars
(Meusinger et al. 1991). Meusinger et al. (1991) derived an
age-metallicity relation similar to that of Twarog (1980b) which
 reflects the preconception that overall metallicity
increases with  time.

What are the biases in the stellar sample defined by Twarog (1980b)?
First the sample is limited in apparent magnitude.   
Both Twarog (1980b) as well
as Meusinger et al. (1991)  corrected for this incompleteness when
calculating the mean [Me/H] for each age bin by weighting the  [Me/H]
for each star by the inverse volume that that particular star would be
visible in given its absolute magnitude. Secondly Twarog (1980b)
found, through simulations, that the cuts in effective temperature
should not influence the derived age-metallicity relation. Our new
investigation into the effects of different cuts in effective
temperature from an sample $\sim$6 times larger than was
available to Twarog shows that indeed the exact cut-off at the cooler
temperature does have large effects on the appearance of the
age-metallicity plot, Fig. \ref{amr_fgk.fig}.  In particular old,
metal-rich stars can be excluded from the sample through the lower cut
in effective temperature.

In Fig. \ref{teff_hist.fig} we show the distributions of  effective
temperatures for the sample studied by Meusinger et
al. (1991). Although the lower cut in effective temperature is low
enough that potential old, metal-rich stars would be included the
number of such stars is very small.  We also compare these
distributions with those of our final sample. 
Also, again, study the plots in Fig.  \ref{amr_fgk.fig} which
shows the effect of excluding, voluntarily or  un-voluntarily, colder
stars. In summary the Twarog (1980b) and Meusinger et al. (1991)
sample suffer from the type of selection effect that we illustrate
in Fig.  \ref{amr_fgk.fig}.

Although the age-metallicity relations derived by Twarog (1980b) and
Meusinger et al. (1991) were successfully corrected for the magnitude
bias it is apparent that if no metal-rich old stars were present in the
original stellar sample then no matter what weighting schemes were
invoked these stars would never be accounted for.

The Edvardsson et al. (1993) sample was, as described in detail in
their article,  chosen in order to study the chemical evolution of the
stellar disk in the Galaxy. The stars were selected from the
$uvby- \beta$ catalogue by Olsen (1983). This catalogue contains nearly
all F stars  brighter than V=8.3. In nine metallicity bins, covering
the metallicities of the galactic disk, 20 slightly evolved dwarf
stars were chosen (for each bin)
for detailed spectroscopic abundance analysis,
giving an even coverage in [Fe/H].  Thus the  metallicity
distribution of that sample is not  representative  for the stellar
populations of the solar neighbourhood as it will necessarily include
more metal-poor stars than are truly present as  compared to
metal-rich. They found an age-metallicity plot which showed an overall
trend such that younger stars were preferentially metal-rich while old
stars were preferentially metal-poor. However, once selection biases
were corrected for the remaining age-metallicity relation is virtually
flat between $\sim 2 - 10$ Gyr and has a $\sigma$ [Fe/H]$ = 0.20$
dex. Perhaps {\sl the} most important point about the Edvardsson et
al. (1993) study is the high degree of internal consistency in  the
derived stellar abundances. In particular the small internal  error on
[Fe/H] which is $\leq 0.05$ dex.  This means that the scatter found in
the age-metallicity plot is  four times larger than the internal error
on [Fe/H] indicating that part of, or all, the scatter is real and not
due to measurement errors, e.g. Nissen (1995). Furthermore, Chen et
al. (2000) applied a similar abundance analysis to 90 F and G dwarf
stars in the solar neighbourhood and found an equally large scatter in
[Fe/H]. For example at 10 Gyr they find a differences in [Fe/H] of 0.8
dex between co-eval stars. In particular they conclude that there
appears to exist metal-rich stars of all ages while metal-poor disk
stars exclusively have old ages.

Ng \& Bertelli (1998) re-derived the stellar ages for the stars in
the Edvardsson et  al. (1993) sample using updated Padua isochrones
(Bertelli et al. 1994) and parallaxes measured by Hipparcos  (ESA
1997). We  have basically used the same method as they did to derive
the stellar ages (see Sect \ref{ages.sect}).  As shown in
Sect. \ref{ng.comp.sect} their ages and ours agree very well.
  Ng \& Bertelli (1998) found
some deviation in derived ages from those presented in Edvardsson et
al. (1993), however, when the Hipparcos parallaxes were relied upon to
determine absolute magnitudes the differences were small. Some
evidence for the revised ages to be larger for old ages and somewhat
younger for young ages was found but the overall impression and trends
seen in the Edvardsson et al.  (1993) plot were recaptured.

Garnett \& Kobulnicky (2000) 
divided the Edvardsson et al. (1993) sample according to distance. In 
their Fig. 2 they found that stars closer then 30 pc presented a 
age-metallicity diagram with considerable scatter while the stars
between 30 and 80 pc appeared to have tighter correlation between
age and metallicity. This effect is not intrinsic to the stars in the
solar neighbourhood but is expected if we consider the way
the Edvardsson et al. sample was constructed. Since they were doing
spectroscopic observations their sample was essentially limited in 
apparent magnitude. That means those stars that they selected that
happened to be further away also are intrinsically more bright than
those that happened to be close by. Consulting Fig. \ref{amr_MV.fig}
we observe that when the intrinsically less bright stars are excluded
then the upper-right hand corner gets depleted, i.e. metal-rich, old
stars are preferentially excluded. Furthermore, Edvardsson et al.
(1993) strived to have equal numbers of stars in each metallicity bin.
As the number density of stars with lower metallicities (i.e. $\sim
-0.5 $ to $ - 1.0$ dex) are smaller than for the more metal-rich stars their
sample naturally included many metal-poor stars that are at a greater
distance from the sun. We can thus see how these two factors conspire
to form a sample where metal-poor, old stars would preferentially
be found in the sub-sample further away from us while the metal-rich 
old stars would be excluded from that sub-sample. Thus the diagrams
presented by Garnett \& Kobulnicky (2000) can naturally be explained 
as part of the sample construction and may not be inherent to the 
stars in general in the solar neighbourhood.

The major conclusion from the Edvardsson et al. (1993) study and the 
Ng \& Bertelli(1998) study was that there does indeed exist a large
spread in the age-metallicity plot. So large that in fact the internal
measurement errors cannot account for it. In particular they found 
that there exist old, metal-rich stars. This is further supported by
Chen et al. (2000) and Feltzing \& Gonzalez (2001), as well as by
the new, larger data sample presented here.

We have already extensively compared our results with those of Rocha-Pinto
et al. (2000). The distribution in $b-y$, from Fig. 1a in Rocha-Pinto et al.
(2000) shows that their sample basically contains stars from 6000 K down to
4500 K, although the sampling becomes progressively patchier at the lower
temperatures. Thus they should not suffer from selection effects in the 
same way as the Twarog sample did and they should in fact have included
old, metal-rich stars. Our main concern with this work and the validity of the 
age-metallicity relation derived stems from our investigation of 
the derivation of chromospheric ages, see Sect. \ref{comp.rochapinto.sect}. 

Several models of Galactic chemical evolution have reproduced the
scatter observed in Edvardsson et al. (1993) and Ng \& Bertelli
(1998), e.g. Pilyugin \& Edmunds (1996), Raiteri et al. (1996),  
and Berczik (1999).  The two latter models take into account not only the
chemical enrichment as time goes but also the dynamical evolution of
the system, while the first model in particular studies the effect of
irregular rates of infall of un-processed  matter onto the stellar
disk. Both approaches create a significant scatter in the
age-metallicity plot supporting our conclusion that
the observed scatter is real and
not a result of observational errors.

\section{Conclusions}

We have presented stellar ages and metallicities for 5828 dwarf and
sub-dwarf stars in the solar neighbourhood. These were used to study
the age-metallicity diagram of the local stellar disk. Because
of the size of the sample, to our best knowledge so far the largest sample
for which such data has been presented, we were able to study, in detail,
the effects of different selection criteria on the age-metallicity
diagram and the conclusions drawn from such diagrams. 

In particular we find that the solar neighbourhood age-metallicity 
diagram is well populated at at all ages and especially that old,
metal-rich stars do exist. Further, that exclusion of colder stars
lead to a loss of old, metal-rich stars. The reason for this can be found
in the stellar evolutionary tracks themselves. Old, metal-rich
stars have to be fairly evolved for stellar isochrones of different
ages to be sufficiently separated in order to allow a reliable
age determination. Thus it is of utmost importance
when selecting a representative stellar sample to include such cold stars.

Further, we found that the scatter in metallicity at any
given age are larger than the observational error, in concordance
with Edvardsson et al. (1993). We also discuss how
 the distance dependence that 
Garnett \& Kobulnicky (2000) found in age-metallicity 
diagram of the Edvardsson et al. (1993) sample
is an  expected effect of the construction of their sample.  

The difference
between our [Me/H] derived from $uvby$ photometry as compared with the
spectroscopic [Fe/H] derived by Edvardsson et al. (1993) is $+0.01\pm 0.10$
dex.

\acknowledgement{Lennart Lindegren is acknowledged for extensive,
fruitful discussions with SF and JH. The authours
would like to thank the referee, W.J. Maciel, for valuable comments and
suggestions that improved the article. Jesper Sommer-Larsen and
Chris Flynn are thanked for providing stimulating critisism of
the text, and   Poul-Erik Nissen and 
Birgitta Nordstr\"om are thanked
for valuable comments and suggestions
on an early draft of the paper. SF thanks The Royal
Swedish Academy of Sciences for financial support during a stay to the
Institute of Astronomy, Cambridge, U.K., 
during which part of this work was completed. JH thanks the
Swedish National  Space Board for financial support. JRH thanks Gerry
Gilmore  and the IoA for support during this work. }

\end{document}